\documentclass[prb,aps,amssym,nofootinbib,floatfix,onecolumn,notitlepage, eqsecnum]{revtex4-1} 
\pdfoutput=1
\usepackage{amsmath,amssymb,bbold,bm}
\usepackage{graphicx}
\usepackage{placeins}
\usepackage{caption}
\usepackage{xcolor}
\usepackage{ulem}

\newcommand{\be}{\begin{equation}}
\newcommand{\ee}{\end{equation}}
\newcommand{\bmx}{\begin{array}}
\newcommand{\emx}{\end{array}}
\newcommand{\bea}{\begin{eqnarray}}
\newcommand{\eea}{\end{eqnarray}}
\newcommand{\dg}{^{\dagger}}
\newcommand{\dn}{^{\vphantom{\dagger}}}

\newcommand{\ra}{\rightarrow}

\newcommand{\bb}[1]{\mathbb{#1}}
\newcommand{\so}{\qquad\rightarrow\qquad}

\newcommand{\pref}[1]{(\ref{#1})}
\newcommand{\tr}[1]{{\rm Tr}\Big[ #1 \Big]}

\newcommand{\diag}[2]{\left(\bmx{cc}#1 & \bb 0 \\ \bb 0 & #2\emx\right)}
\newcommand{\abs}[1]{\left\vert #1 \right\vert}
\newcommand{\mat}[1]{\left(\bmx{cc}#1\emx\right)}
\newcommand{\diagn}[2]{\bmx{cc}#1 & \bb 0 \\ \bb 0 & #2\emx}

\newcommand{\bw}[1]{\begin{widetext}}
\newcommand{\ew}[1]{\end{widetext}}

\begin{document}
\title{Effect of Disorder on the Conductance of (non-) Topological SN Junctions}
\author{Yashar Komijani$^{1,2}$}\email{komijani@phas.ubc.ca}
\author{Ian Affleck$^1$}
\affiliation{ $^1$Department of Physics and Astronomy and $^2$Quantum Materials Institute, University of British 
Columbia, Vancouver, B.C., Canada, V6T 1Z1}
\date{\today}
\begin{abstract}
General multi-channel SN junctions fall into two topological classes depending on whether or 
not there is a Majorana mode localized at the junction. This is known  to lead to different behaviour of 
the conductance  in the presence of arbitrary disorder 
near the junction.  We discuss these topological properties from two perspectives, one based on 
representing the disorder by a scattering matrix in series with that of a clean SN junction
and one based on low energy field theory methods. The first approach 
is used to discuss the effect of an ohmic contact between a quantum wire and a three dimensional metal 
far from the junction. The second is useful for treating interactions. 
\end{abstract}
\maketitle
\section{Introduction}

It has been recently predicted that superconductor-normal (SN) junctions in quantum wires with spin-orbit coupling can host a localized 
Majorana mode (MM) and that this leads to a  linear conductance of $G=2e^2/h$ at zero temperature [\onlinecite{Kitaev01,Oreg10,Lutchyn10,Alicea12,Mourik12,Das12,Deng12}] in a simplified model containing only one tranverse subband. 
More realistic models of quantum wires will have more partially occupied transverse subbands, with Hamiltonian of the form:
\be
H=\int {d^3x}\{\psi\dg (\vec x)\Big[{(\vec p)^2\over 2m}-\mu+V(\vec x)+\alpha (p_x \sigma^y-p_y\sigma^x)+B(\vec x)\sigma^x\Big]\psi (\vec x)
+\Big(\Delta (\vec x)\psi_\uparrow (\vec x)\psi_\downarrow (\vec x)+h.c.\Big)\Big\}+H_{int}.\label{Hq}
\ee
The quantum wire runs in the $x$-direction, with a small finite width in the $y$ and $z$ directions allowing 
for several occupied transverse subbands. $\alpha$ is the Rashba spin-orbit interaction coefficient. The proximity-effect induced superconducting gap is non-zero for $x<0$, turning 
off smoothly near $x=0$. 
$V(\vec x)$ represents a combination of gate voltages and disorder.
$B(\vec x)$ is a Zeeman magnetic field pointing in the $z$ direction which might also vary spatially. We assume the magnetic field is small enough (compared to  the quantum wire width) that orbital effects can be neglected.
(We set $g\mu_B=2$.) 
If $B(\vec x)$ is sufficiently large in the superconducting region then the system will be in the topological phase, 
with a Majorana mode localized near $x=0$. 
 We focus here on the effects of disorder 
on the normal side of the junction, assuming the superconducting side is sufficiently clean to be 
everywhere in the topological phase. 
We may decompose $\psi (\vec x)$ into several transverse subbands each of which has 2 spin components. 
We label the total number of active channels (including a factor of $2$ for spin) as $N$.
 $N$ may be odd or even depending on whether the Fermi energy is or is not between the minimum energies of the 
highest spin-split channels [\onlinecite{odd}].

Depending on details all of these channels may couple to the MM. We do not assume any particular symmetry of this Hamiltonian which puts it in the class D of the so-called tenfold-way symmetry classes [\onlinecite{Altland97},\onlinecite{Ryu10}] (see Appendix A for more discussion on symmetry).
General results were derived for the zero bias, zero temperature conductance of such 
a system using random matrix theory. The topological and non-topological cases are distinguished simply by whether the 
reflection matrix has determinant $Q=-1$ or $1$  respectively. 
The probability distribution of the conductance for chaotic scattering was analysed; it can take any value between 
lower and upper bounds which were determined as a function of $N$ and $Q$ [\onlinecite{Beenakker11,Diez,Beenakker14}]. 

Here we analyze the topological properties of such a junction from two perspectives. 
One is based on considering a clean SN junction, represented by a reflection matrix $\bb r$, in series 
with a scattering region on the normal side corresponding to disorder 
and represented by an S-matrix $S_1$. $\bb r$ represents both normal and Andreev reflection 
and its determant is $Q=\pm 1$ for a non-topological or topological junction respectively. While $S_1$ 
may contain Andreev as well as normal reflection and transmission, we assume it has determinant $+1$. 
We  show that the total reflection matrix for the combined scattering system has determinant $Q$.  
We then use this approach to analyze an SN junction in a dirty quantum wire with an ohmic contact to 
a three dimensional (3D) metal, far from the junction. We derive upper and lower bounds on the conductance 
in this case, proving that they are determined by the number of channels in the quantum wire, and $Q$, only, 
independent of properties of the 3D metal. Our second approach is based on a low energy effective 
relativistic field theory, valid at energy scales low compared to the superconducting gap 
and also compared to $v_F/\ell$ where $\ell$ is the length 
of the disordered region on the normal side near the junction. Then we can integrate out 
all degrees of freedom near the junction except for the Majorana mode. This integrating out procedure generates  scattering terms in the effective Hamiltonian, localized at the junction, which 
represent the disorder. We show 
that these scattering terms can be eliminated from the effective Hamiltonian by a unitary transformation
which does not change the sign of the determinant ($-1$) resulting from the Majorana mode. 
This field theory approach is useful in treating interactions in the normal part of the wire [\onlinecite{KAint}].

In the next section we review the topological classification of SN junctions and the 
resulting bounds on the conductance. In Sec. III we discuss our series treatment of 
disorder and study the effects of an ohmic contact.  Sec. IV contains our relativistic 
field theory treatment. Technical details are given in two Appendices, which include 
an alternative derivation of the conductance bounds.

\section{Topology of the S-matrix and conductance : review}
The total $T=0$ linear conductance (summed over all channels) of an $N$ channel SN junction can be written [\onlinecite{BTK}]:
\be G={e^2\over h}\sum_{i=1}^N\left[1-\sum_{j=1}^N |{\bb r}^{ee}_{ij}|^2+\sum_{j=1}^N |{\bb r}^{eh}_{ij}|^2\right].\label{BTK}\ee
Here ${\bb r}_{ij}^{ee}$ is the amplitude for an incoming electron in channel $j$ to be reflected as an electron in channel $i$ 
and ${\bb r}^{eh}_{ij}$ is the amplitude to be reflected as a hole. The reflection amplitudes are calculated at zero energy. For their 
precise definition (into which a factor of the square root of the ratio of Fermi velocities in channels $i$ and $j$ has been adsorbed) see
 Appendix A. This formula, due to Blonder, Tinkham and Klapwijk [\onlinecite{BTK}], has a simple Landauer-like interpretation. The first term inside the brackets 
represents the current due to the incoming electrons in channel $i$; the second and third terms represent the current due to the 
reflected particles and holes. Due to the superconducting gap, there is no quasi-particle current at $T=0$ and zero source-drain voltage 
inside the superconductor; the last term in Eq.\,(\ref{BTK}) represents Cooper pairs being transmitted into the superconductor 
during Andreev reflection. It is convenient to assemble normal and Andreev reflection amplitudes into a $2N\times 2N$ matrix. 
\be {\bb r}=\left(\begin{array}{cc}
{\bb r}^{ee}& {\bb r}^{eh}\\
{\bb r}^{he}&{\bb r}^{hh}
\end{array}\right).\ee
Conservation of quasi-particle current implies that ${\bb r}$ is unitary.
\be {\bb r}^\dagger={\bb r}\label{run}.\ee
  Furthermore, the electron-hole symmetry of the Bogliubov de Gennes equation (Appendix A)
implies, at zero energy:
\be {\bb r}^{ee}={\bb r}^{*hh},\ \  {\bb r}^{eh}={\bb r}^{*he}\ee
or equivalently
\be {\bb r}=\tau^x{\bb r}^*\tau^x.\label{reh}\ee
Eqs.\,(\ref{run}) and (\ref{reh}) imply that $ {\bb r}$ is unitarily equivalent to a real orthogonal matrix $ {\bb r}_M$:
\be {\bb r}\equiv {\bb C}{\bb r}_M\bb{C}^\dagger\label{rM}\ee
where
\be
\bb {C}\equiv {1\over \sqrt{2}}\left(\begin{array}{cc} \bb{1}&i\bb{1}\\
\bb{1}&-i\bb{1} \end{array}\right).\label{Cdef}
\ee
Here $\bb{1}$ is the $N\times N$ unit matrix.  
We use the label $M$ for the orthogonal matrix because $\bb {C}$ transforms the fermion fields on the normal side 
to the ``Majorana basis'' of Hermitian operators, as discussed in Sec. IV.  The conductance formula of Eq.\,(\ref{BTK}) 
can be written in terms of the orthogonal matrix $\bb {r}_M$ [\onlinecite{Pikulin13}]:
\be
G=\frac{e^2}{h}\left[N-\frac{1}{2}\hbox{Tr}\Big({\bb r}^\dagger \tau^z {\bb r}\tau^z\Big) \right]=\frac{e^2}{h}\left[N-\frac{1}{2}\hbox{Tr}\left({\bb r}_M^T \tau^y{\bb r}\dn_M\tau^y\right)\right].
\label{GM}\ee
Here $\tau^z$ and $\tau^y$ are the Pauli matrices acting in the electron-hole space:
\be \tau^y\equiv \left(\begin{array}{cc}0&-i\bb{1}\\i\bb{1}&0\end{array}\right).\ee
A crucial observation is that the 
set of orthogonal matrices breaks up into 2 topological classes with det ${\bb r}_M=\pm 1$. The sign of the determinant is completely 
determined by whether the SN junction is topological or not [\onlinecite{Merz02,Pikulin11,Fulga11,Akhmerov11,Fulga12}]:
\bea \det{\bb r}_M&=&1,\  \  (\hbox{non-topological junction})\nonumber \\
&=&-1,\  \  (\hbox{topological junction}).\label{1-1}\eea
This can be seen by considering simple examples. 
Perfect normal reflection corresponds to ${\bb r}_M=\bb{1}$ and hence det ${\bb r}_M=1$.  On the other hand, consider a simple clean topological SN junction which has 
 only the first channel coupled to the MM resulting in perfect Andreev reflection with the other channels decoupled, having 
normal reflection, diagonal in the channel index. This corresponds to 
\bea {\bb r}^{eh}_{ij}&=&\delta_{i1}\delta_{j1}\nonumber \\
{\bb r}^{ee}_{ij}&=&\delta_{ij}(1-\delta_{i1})\eea
and hence
\be {\bb r}_M=\left(\begin{array}{cccccc}1&0&0&0&\ldots &0\\
0&-1&0&0&\ldots &0\\
0&0&1&0&\ldots &0\\
0&0&0&1&\ldots &0\\
\vdots&\vdots&\vdots&\vdots&\ddots&\vdots\\
0&0&0&\ldots&\ldots&1
\end{array}\right)\ee
with determinant $-1$.  We may now complicate the Hamiltonian and hence the reflection matrix by mixing channels, adding disorder, et cetera. Any continuous change 
in the Hamiltonian cannot lead to a discontinuous jump in det ${\bb r}_M$ leading to Eq.\,(\ref{1-1}). Further evidence for this is provided in Secs. III and IV.

\begin{figure}[h!]
\includegraphics[width=.5\linewidth]{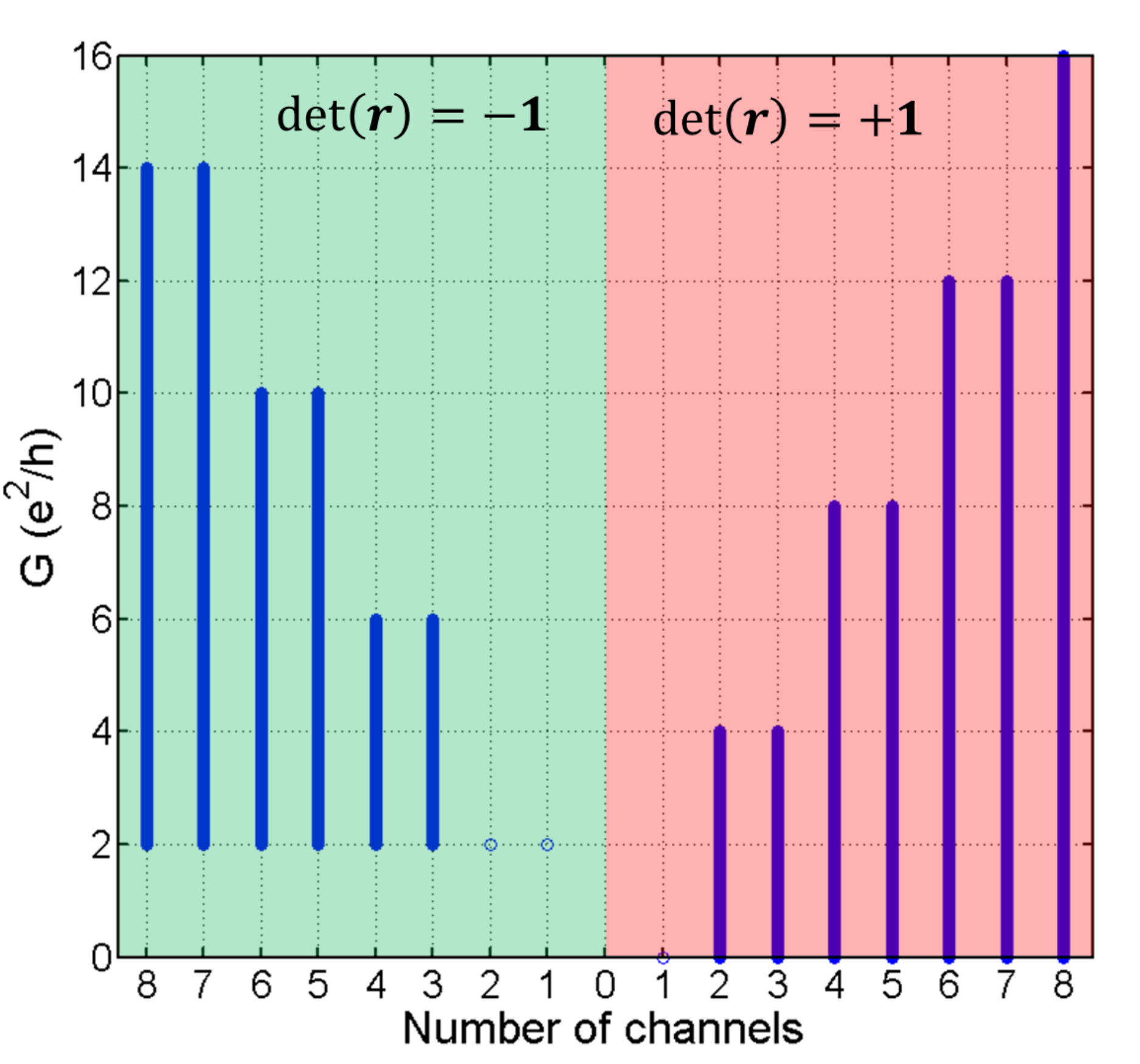}
\caption{\small\raggedright (color online) The range of the conductance vs. the number of spin-resolved channels, from Eq.\,(\ref{Gb}) for the two cases of $\det{r}=\pm 1$. The single-channel case in the non-topological regime ($\det(r)=1$) and single- and double-channel cases in the topological regime ($\det(r)=-1$) are special in the sense that the conductance is universal and robust against disorder.
\label{fig:Grand}}
\end{figure}

For the case of $N=1$, particle-hole symmetry leads to $G=(1-\det{\bb r_M})e^2/h$ and the conductance is only determined by the topology. A generalization of this formula to higher $N$ was provided in [\onlinecite{Diez}] for class BDI junctions and in [\onlinecite{Beenakker11}] for class D junctions using polar decomposition of the $\bb r$ matrix and the so-called B\'eri degeneracy [\onlinecite{Beri}] of the eigenvalues of $\bb r^{he}$. The conductance for different channels and different topologies are given by
\bea 0&\leq& {h\over e^2}G \leq 2N,\ \  (N\ \hbox{even},\ \hbox{det} \ {\bb r}_M=1)\nonumber \\
 0&\leq& {h\over e^2}G \leq 2N-2,\ \  (N\ \hbox{odd},\ \hbox{det} \ {\bb r}_M=1)\nonumber \\
 2&\leq& {h\over e^2}G \leq 2N,\ \  (N\ \hbox{odd},\ \hbox{det} \ {\bb r}_M=-1)\nonumber \\
   2&\leq&{h\over e^2}G \leq 2N-2,\ \  (N\ \hbox{even},\ \hbox{det} \ {\bb r}_M=-1).\label{Gb2}
\eea
These ranges are plotted in Fig.\,  (\ref{fig:Grand}).  
An alternative derivation of these ranges is presented in Appendix B. Apart from $N=1$ case discussed above, we see that $N=2$ topological case is also interesting as the conductance is uniquely determined  by the topology, $G=2e^2/h$.
Note that, in general for a topological SN junction these ranges imply $G>2e^2/h$. A simple situation in which this occurs is 
when the various transverse subbands are unmixed, with only one of them in the topological phase. Then the conductance is 
simply the sum of $2e^2/h$ from the topological subband plus the contributions from all the non-topologicial subband. 
Without magnetic field or SOI, for small $|\Delta |$, and approximating the disordered junction 
as an ideal SN junction in series with a normal scattering region, the contribution from the $(N-2)/2$ non-topologicial subbands would be
 [\onlinecite{NazarovBlanter}]
\be G_{\hbox{non-topological}}={e^2\over h}\sum_{i=1}^{(N-2)/2}\frac{T_i^2}{(1-T_i/2)^2}.\label{Gdec} \ee
where $T_i$ is the normal transmission probability through the junction for the $i^{\rm th}$ non-topologicial subband. 
We expect this formula to remain true in the presence of spin orbit interaction and Zeeman field which have 
small characteristic energies compared to the Fermi energy and band width. We thus obtain
\be G\approx {e^2\over h}\left[2+\sum_{i=1}^{(N-2)/2}\frac{T_i^2}{(1-T_i/2)^2}\right] >{2e^2\over h}.\label{Gdec2} \ee

\FloatBarrier
\section{Two Scattering Regions in Series}\label{sec:bulk}
\begin{figure}[h!]
\includegraphics[width=0.4\linewidth]{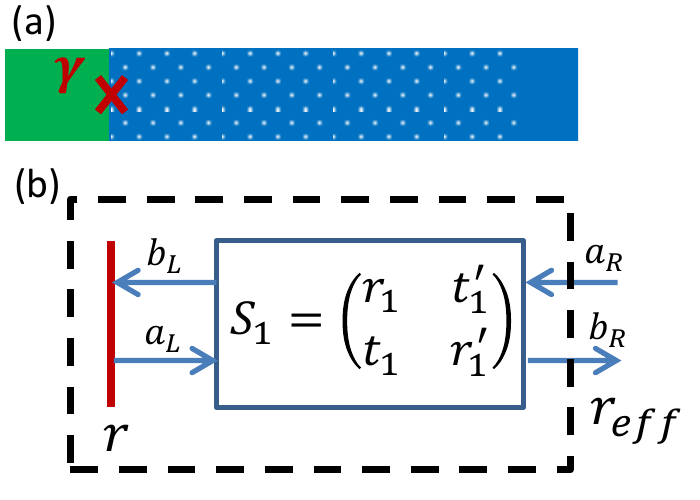}
\caption{\small\raggedright (a) Schematic of a topological SN interface with a Majorana mode $\gamma$ between the superconductor (green) and disordered normal wire.\,(b) The combination of the SN interface and the disordered wire can be modeled by an effective reflection matrix $\bb{r}_{eff}$, composed of the matrix $S_1$ describing the normal wire and matrix $\bb{r}$ describing the MM.\label{fig:cascade}}
\end{figure}

Another way of understanding the topological invariance of the determinant of the S-matrix in a system with disorder near the SN junction is to represent 
the disordered part of the normal wire by an $S$-matrix, $S_1$, as sketched in Fig.\,(\ref{fig:cascade}). This implies that $S_1$ is a $4N\times 4N$ unitary matrix allowing for incoming 
particles or holes from the left or right side in any of $N$ channels. $S_1$ relates incoming and outgoing waves to the left and right of the scattering region:
\be
\mat{\vec b_L\\\vec b_R}=S_1\mat{\vec a_L\\\vec a_R},
\qquad
S_1=\mat{\bb r_1 & \bb t'_1 \\ \bb t_1 & \bb r'_1}.\label{eqC1}
\ee
To be able to define this matrix, the normal wire has to be clean beyond a certain distance from the SN junction.
 Here $\vec a_{L/R}$ and $\vec b_{L/R}$ denote incoming and outgoing $2N-$dimensional vectors of amplitudes from left/right side, respectively.
The total reflection matrix is then obtained by combining the reflection matrix $\bb r$ of 
the clean SN junction with $S_1$. It would be natural to assume that $S_1$ only contains normal reflection and transmission amplitudes 
however, our proof continues to work even if $S_1$ also contains Andreev processes, provided that $\det(S_1)=1$, which should 
be true provided there is no unpaired Majorana mode localized in the disordered part of the normal wire represented by $S_1$. This can be seen especially in the opaque regime in which $\det(S_1)\ra\det(\bb r_1)\det(\bb r'_1)$.
The effective reflection matrix for a disordered SN junction is obtained by combining this matrix with the reflection matrix of the clean SN junction, which satisfies $\vec a_L=\bb{r}\vec b_L$. This equation can be used to eliminate $\vec a_L$ and $\vec b_L$ from Eq.\,\pref{eqC1} and show that the series system obeys $\vec b_R=\mathbb{r}_{eff}\vec{a}_R$ with an effective reflection matrix (of dimension  $2N\times 2N$)
\be
\bb{r}_{eff}=\bb{r}'_{1}+\bb{t}_{1}(\bb{1}-\bb{r}\bb{r}_{1})^{-1}\bb{r}\bb{t}'_{1}.\label{eqdisorder1}
\ee
It can be easily seen that $\bb{r}_{eff}$ is unitary and  particle-hole symmetric if $\bb r$ and $S_1$ are. One way to see this is to use the Majorana representation of these matrices, Eq.\,(\ref{rM}) in which they are real and orthogonal, and show that $\bb r_{eff}$ defined by Eq.\,\pref{eqdisorder1} is also real and orthogonal (see Appendix C). We now prove the important property
\be
\det(\bb r_{eff})=\det(\bb r)\det(S_1).\label{eqdisorder2}
\ee
It is convenient to form a new $4N\times 4N$ matrix $S_2$, out of the reflection matrix $\bb{r}$ and the scattering matrix $S_1$ by
\be
S_2\equiv\mat{\bb r_2 & \bb t'_2 \\ \bb t_2 & \bb r'_2}\equiv\mat{\bb r & \mathbb{0} \\ \mathbb{0} & \bb r} S_1.\label{eqS2}
\ee
Obviously $S_{2}$ is orthogonal and we have $\det(S_{2})=\det(S_1)$. Next, we start with the identity
\bea
\mat{\bb r_2-\bb{1} & \bb t'_2 \\ \bb t_2 & \bb r'_2}&=&\mat{\bb{r}_2-\bb{1} & \bb 0 \\ \bb{t}_2 & \mathbb{1}}\mat{\bb{1} & (\bb r_2-\bb{1})^{-1}\bb{t}_2' \\ \bb{0} & \bb r'_2-\bb t_2(\bb r_2-\mathbb{1})^{-1}\bb t'_2}\nonumber
\eea
Here $\mathbb{1}$ and $\mathbb{0}$ represent corresponding diagonal $2N\times 2N$ matrices. Taking determinant of both sides we get
\be
\det(S_2-P_L)=\det(\bb r_2-\mathbb{1})\det(\bb r\bb r_{eff}).
\label{eq5}
\ee
Here $P_L\equiv\rm{diag}(\bb{1},\bb{0})$. Now we use the fact that $S_{2}$ is an orthogonal matrix and write
\bea
\det(S_2-P_L)&=&\det(S_2)\det(\mathbb{1}-S_2^T P_L)\\
&=&\det(S_{1})\det(\mathbb{1}-\bb r_2)
\eea
Using the fact that $\bb r_2-\mathbb{1}$ has even dimension, we have $\det(\bb 1-\bb r_2)=\det(\bb r_2-\bb 1)$. 
This together with Eq.\,\pref{eq5} proves that $\det(S_{1})=\det(\bb r_{eff})\det(\bb r)$ which eventually proves the desired Eq.\,\pref{eqdisorder2}. Note that we did not make any assumption about the matrices except that they are orthogonal (unitary and obey particle-hole symmetry) and this derivation is valid for arbitrary $N$.  Thus we conclude that the additional scattering from disorder, corresponding to $S_1$, 
does not change sign of the determinant of the reflection matrix. Whether the junction is topological (det ${\bb r}_{eff}=-1$) or non-topological (det ${\bb r}_{eff}=1$) 
is unaffected by disorder [\onlinecite{Brouwer11}, \onlinecite{Wimmer11}].

\subsection{Treating contacts using series approach}
\begin{figure}[h!]
\includegraphics[scale=1]{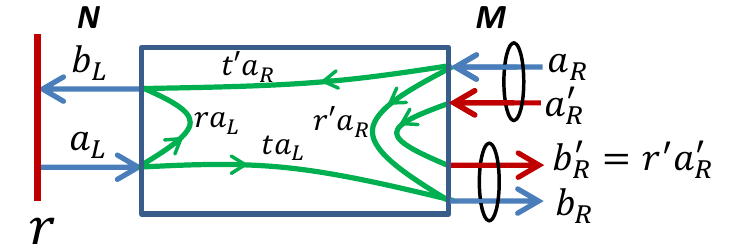}
\caption{\small\raggedright The box represents a normal contact between the $N$ channel quantum wire on the left and an $M$ channel 
3D metal on the right. Only $N$ of the $M$ channels, $a_R$, are transmitted through the contact, the other
$M-N$, $a_R'$ being totally reflected. The partially reflected channels $a_R$ reflect into outgoing channels, $b_R$,  
linearly independent of the outgoing totally reflected channels $b_R'$. The channels, $a_L$ transmitted 
through the contact end up purely in the $b_R$ subspace.}
\label{fig:contact}
\end{figure}

An interesting thing about our series approach is that Eq.\,(\ref{eqdisorder1})
remains valid when the left and right sides have different number of channels, $N$ on the left side 
and $M$ on the right side as sketched in Fig.\,(\ref{fig:contact}). To see this, note that the matrix $S_1$ can be a $2(N+M)\times 2(N+M)$ matrix, composed of $2N\times 2N$ matrix $\bb{r}_1$, $2N\times 2M$ matrix $\bb t'_1$, $2M\times 2N$ matrix $\bb t_1$ and $2M\times 2M$ matrix $\bb r'_1$. $S_1$ is still unitary $S_1\dg S\dn_1=\bb 1$ and $S_1\dn S_1\dg=\bb 1$ and it obeys particle-hole symmetry $S\dn_1=\tau^xS_1^*\tau^x$. We can again write $\vec a_L=\bb{r}\vec b_L$ where $\bb{r}$ is $2N\times 2N$ and after eliminating $\vec a_L$ and $\vec b_L$, we see that $\vec a_R$ and $\vec b_R$ obey $\vec b_R=\mathbb{r}_{eff}\vec{a}_R$ with $2M\times 2M$ matrix $\bb r_{eff}$ given above. Interestingly, even the determinant formula carries over to this case, i.e. $\det\bb r_{eff}=\det\bb r\det S_1$.

Here, we would like to use these properties, to discuss a contact between the disordered nanowire with $N$ channels (coupled to the MM) and a 3 dimensional metal characterized by $M\gg N$ channels. After combining the reflection matrix of the disordered nanowire $\bb r$ with the S-matrix of the ohmic contact $S_1$ we obtain a $2M\times 2M$ matrix $\bb r_{eff}$. The question is what is the maximum conductance that this system can exhibit? Is it given by the  number of channels $N$ in the quantum wire 
or $M$ in the 3D metal? See
Fig.\,(\ref{fig:contact}).

To answer this, we assume that the  contact is far enough from the MM and the SN junction, (in a clean nanowire, the relevant characteristic length scales are Majorana screening cloud [\onlinecite{AG2}] $v_F^2/t^2$ and the coherence length of the superconductor $v_F/\Delta$), that there are no Andreev processes taking place at the interface. Therefore, we can assume that the $S_1$ is block diagonal in electron and hole sectors, $S_1={\rm diag}(S_c,S_c^*)$. Here $S_c$ is an $(N+M)\times (N+M)$ matrix.
\be
\mat{\vec b_L \\ \vec b_R}=S_c\mat{\vec a_L \\ \vec a_R}, \qquad S_c=\mat{\bb r_c & \bb t'_c \\ \bb t_c & \bb r'_c}
\ee
First, we look for solutions in which, an incoming $\vec a'_R$ is totally reflected and does not produce a $\vec b_L$. In other words
\be
\bb 0=\bb t'_c\vec a'_R
\ee
This contains $N$ equations for the $M$ unknowns $a_{Ri}'$ and gives $M-N$ linearly independent incoming waves on the 
 right side that are totally reflected and only $N$ channels, $\vec a_R$ which are (partially) transmitted 
to the left side: $\bb t'_c\vec a_R\neq 0$.

Next we note that the fully reflected states, ${\bb r}'_c\vec a_R'$, are orthogonal both 
to the reflected part of the $\vec a_R$ 
states, ${\bb r}_c'\vec a_R$  and also to the  states transmitted from the left side, ${\bb t}_c\vec a_L$.
This follows simply because
\bea
(\bb r'_c\vec a_R)^*\cdot (\bb r'_c\vec a'_R)=\vec a_R\dg(\bb r'_c{\dg}\bb r'_c)\vec a'_R=\vec a_R\dg(\bb 1-\bb t_c'{\dg}\bb t'_c)\vec a'_R=0 \so \bb r'_c\vec a_R \perp \bb r'_c\vec a'_R
\\
(\bb t_c\vec a_L)^*\cdot (\bb r'_c\vec a'_R)=\vec a_L\dg(\bb t_c\dg\bb r'_c)\vec a'_R=\vec a_L\dg(-\bb r\dg_c\bb t'_c)\vec a'_R=0 \so \bb t_c\vec a_L \perp \bb r'_c\vec a'_R.
\eea
For the second equalities in both lines we have used unitarity of the $S_c$ matrix implying:
\bea {\bb t}_c^\dagger {\bb r}_c'+{\bb r}_c^\dagger {\bb t}_c'&=&0\nonumber \\
 {\bb r}_c'^\dagger {\bb r}_c'+{\bb t}_c'^\dagger {\bb t}_c'&=&\bb 1.\eea
Thus we see that the fully reflecting states, $\vec a_R'$, $\vec b_R'$, completely decouple from the partially transmitting states, $\vec a_R$, $\vec b_R$, 
$\vec a_L$, $\vec b_L$. This implies that unitary transformations:
\be \vec a_R\to U_a\vec a_R.\ \   \vec b_R\to U_b\vec b_R\ee
transform  $S_c$ so that $\bb r'_c$ transforms to a propagating block $\tilde{\bb r}'_{cp}$ and a reflected part $\tilde{\bb r}'_{cr}$. Therefore, we have
\be 
U_b\dg\bb t_c=\mat{\tilde{\bb t}_c \\\hline \bb{0}}, \qquad \bb t'_c U_a\dn=\mat{\tilde{\bb t}'_c \ \ \vline& \bb 0}, \qquad
U_b\dg\bb r'_cU_a\dn=\left(\bmx{c|c}\tilde{\bb{r}}'_{cp} & \bb 0 \\ \hline \bb 0 & \tilde{\bb r}'_{cr}\emx\right)\equiv \tilde{\bb r}'_c.
\ee
The horizontal/vertical line separate the first $N$ row/columns from the other $M-N$ rows/columns. $\tilde{\bb t}_c$, $\tilde{\bb t}'_c$ and $\tilde{\bb r}'_{cp}$ are $N\times N$ matrices and $\tilde{\bb r}'_{cr}$ is an $(M-N)\times(M-N)$ matrix.

Extending these unitary transformations to the hole sector by $U_{A,B}\equiv{\rm diag}(U\dn_{a,b},U_{a,b}^*)$ and applying them to 
$\bb r_{eff}$ in the particle-hole space we obtain
\bea
U_B\dg\bb r_{eff} U_A\dn&=&U_B\dg\diag{\bb r'_c}{\bb r_c'^*}U_A+U_B\dg\diag{ \bb t_c}{\bb t_c^*}\Bigg[\bb 1-\bb r \diag{\bb r_c}{\bb r_c^*}\Bigg]^{-1}\bb r\diag{\bb t'_c}{\bb t_c'^*} U_A\\
&=&
\left(\bmx{c|c}\diagn{\tilde{\bb r}'_{cp}}{\tilde{\bb r}_{cp}'^*} & \diagn{\bb 0}{\bb 0} \\ \hline \diagn{\bb 0}{\bb 0} & \diagn{\tilde{\bb r}'_{cr}}{\tilde{\bb r}_{cr}'^*}\emx\right)
+\left(\bmx{c}\diagn{\tilde{\bb t}_c}{\tilde{\bb t}_c^*} \\
\hline \diagn{\bb 0}{\bb 0}
\emx\right)
\Bigg[\bb 1-\bb r 
\diag{\bb r_c}{\bb r_c^*}
\Bigg]^{-1}\bb r
\left(\bmx{c|c}\diagn{\tilde{\bb t}'_{c}}{\tilde{\bb t}_{c}'^*} & \diagn{\bb 0}{\bb 0}\emx\right)
\eea
where the lines separate the first $2N$ components and the second $2(M-N)$ components. The only off-diagonal elements in electron-hole basis come from $\bb r$. Defining $\tilde{\bb t}_1={\rm diag}({\tilde{\bb t}\dn_c},{\tilde{\bb t}^*_c})$ and similar ones for $\tilde{\bb r}'_{1p}$, $\tilde{\bb r}'_{1r}$, $\tilde{\bb t}'_{1}$ and $\tilde{\bb r}_{1}$ we can write the transformed $\bb r_{eff}$ in channel space as
\be
\tilde{\bb r}_{eff}=U_B\dg \bb r_{eff} U_A=
\left(\bmx{c|c}\tilde{\bb r}'_{1p}+\tilde{\bb t}_1(1-{\bb r} \tilde{\bb r}_1)^{-1}{\bb r}\tilde{\bb t}'_1 & \bb 0 \\ \hline
\bb 0 & \tilde{\bb r}'_{1r}\emx\right)
\ee
where $\tilde{\bb r}'_{1r}$ is block-diagonal in electron-hole space.
An important feature of the conductance formula
\be
G_M=\frac{e^2}{h}\Big(M-\frac{1}{2}\tr{\tau^z\bb r\dg \bb\tau^z \bb r\dn}\Big)
\ee
is that for block-diagonal $\bb r$ matrices, it is additive because $\tau^z$ only acts in 
the electron-hole subspace. In other words, we can write $G_M=G_N+G_{M-N}$. $G_N$ is given by the first block of the $2N\times 2N$ matrix $\tilde{\bb r}_{eff}$. The other contribution $G_{M-N}$ is composed of the second block of the $2(M-N)\times 2(M-N)$ matrix $\tilde{\bb r}_{1r}$ of $\tilde{\bb r}_{eff}$ which is totally diagonal in electron-hole space. Thus it commutes with $\tau^z$, and since it is unitary, we get $G_{N-M}=0$. Therefore, the conductance is completely determined by $N$ linear combination out of $M$ channels and $G_M=G_N$.
Thus for arbitrary disorder in the quantum wire and at the contact, the conductance bounds corresponding to 
$N$ channels apply, provided the conductance is sufficiently far from the junction. 
\section{Relativistic Field theory treatment}
In order to understand the low energy conductance of this system, and, in particular, to include the effects of interactions in the 
normal portion of the wire [\onlinecite{KAint}], it is convenient [\onlinecite{ACZ,Fidkowski12,AG}] to write a low energy effective Hamiltonian, valid at energy scales $\ll \Delta$, 
the induced superconducting gap. 
All degrees of freedom in the superconducting portion of the wire 
are integrated out in a Feynman path integral approach, leaving only the MM. The dispersion relations are linearized 
for the $N$ channels in the normal region.  The effective Hamiltonian, written in the $x\geq 0$ region only, is then 
a sum of  the bulk $H_0$ term for the normal wire,  a boundary term $H_b$ containing 
the tunnelling between the MM and the $N$ normal channels and $H_d$ including the effects of disorder near the junction:
\bea 
H&=&H_0+H_b+H_d\vphantom{\Big] }\nonumber \\
H_0&=&i\int_0^{\infty} dx \sum_{i=1}^Nv_{Fi}\left[\psi_{Ri}^\dagger {d\over dx}\psi_{Ri}-\psi_{Li}^\dagger {d\over dx}\psi_{Li}\right]\nonumber \\
H_b&=&\gamma \sum_{j=1}^Nt_j[\psi_j(0)-\psi_j^\dagger (0)].\label{Hcon}
\eea
Here $L$, $R$ label left and right movers and a boundary condition is imposed, $\psi_{Rj}(0)=\psi_{Lj}(0)\equiv \psi_i(0)$. 
$j$ labels the two spin split bands with different Fermi velocities, $v_{Fi}$. $\gamma$ is the MM operator, obeying $\gamma^\dagger =\gamma$, $\gamma^2=1$. $H_d$ represents an additional set of   boundary interactions at $x=0$, given by
\be H_d=\psi^\dagger (0)\bb M\psi (0)+
[\Delta_b\psi_\uparrow (0)\psi_\downarrow (0)+h.c.]\label{Hd}\ee
where a sum over channel indices is implied in the first term and 
the $N\times N$ matrix $\bb M$ is Hermitian. Assuming that the disorder is extended over a finite length $\ell$ from the interface, at energies smaller than $\bar v/\ell$, it can be absorbed by the boundary interactions given above. These can be eliminated from the low energy Hamiltonian by a unitary 
transformation to the scattering basis (sec.\,\ref{sec:bndary}).

\subsection{Channel-Resolved Conductance}
Here we temporarily ignore $H_d$. 
Despite the different Fermi velocities, this model actually has an $SU(N)$ symmetry before turning on the tunnelling terms, $t_j$, 
upon defining rescaled fields so as to preserve the canonical anti-commutation relations:
\be \psi_i'(x)\equiv \left({v_i\over \bar v}\right)^{1/2}\psi_i\left({v_i\over \bar v}x\right)\label{psi'}\ee
where $\bar v$ is an arbitrary velocity scale which drops out of physical quantities. It is then convenient to make an orthogonal transformation to a new basis of channels, 
\be \tilde \psi_1\equiv \frac{t_1'\psi_1'+t_2'\psi_2'+\cdots+t_N'\psi'_N}{\sqrt{\sum_{j=1}^Nt_j^{'2}}},\qquad \tilde \psi_2\equiv \frac{t_2'\psi_1'-t_1'\psi_2'+\cdots+t_N'\psi'_{N-1}-t_{N-1}'\psi'_N}{\sqrt{\sum_{j=1}^Nt_j^{'2}}},\qquad\cdots\ee
(where $t_i'\equiv t_i\sqrt{\bar v/v_i}$) 
such that only $\tilde \psi_1$ couples to the MM. A single normal channel coupled to a MM is known to exhibit 
perfect Andreev reflection at zero energy [\onlinecite{LawLeeNg09}, \onlinecite{Flensberg10}].
This is because the coupling to MM is infrared relevant in a renormalization group sense, so it grows at low energies and tends to enforce the boundary condition
 $\tilde\psi\dn_1(0)=\tilde\psi_1\dg(0)$, the signature of pure Andreev reflection. Assuming that all the other channels are normally reflected, they do not  contribute to the conductance. It then follows that, at zero energy, the channel-resolved linear conductances of the $N$ channels are
\be G_i={2e^2\over h}{(t_i')^2\over\sum_{j=1}^N( t_j')^2}\ee
with total conductance, $G\equiv G_1+G_2+\cdots+G_N=2e^2/h$. Thus, a convenient way to determine the tunnelling parameters $t_j$ in the 
low energy effective Hamiltonian is by measuring the conductances $G_i$ for a given microscopic model. Although channel-resolved conductance is not topological, the easier-to-measure total conductance has some topological relevance as we saw above.
\subsection{Disorder and boundary interactions}\label{sec:bndary}
Here we consider the effects of the additional  boundary interactions of Eq.\,\pref{Hd}, which represent the effects of disorder near the SN interface 
in the low energy effective Hamiltonian approach. 
It is convenient to make an ``unfolding transformation'',  writing $H_0$ in terms of left-movers only 
on the infinite line, $-\infty <x<\infty$, by defining:
\be \psi_L(-x)\equiv \psi_R(x),\ \  (x>0).\ee
We then make the change of basis in Eq.\,(\ref{psi'}) and
finally, we go over to a basis of Majorana fermions
defining:
\be 
\mat{\psi_j'(x)\\ \psi_j'^\dagger (x)}={\bb C\over 2}
\mat{\gamma_{2j-1}(x)\\ \gamma_{2j}(x)}\label{Majbasis}
\ee
where the matrix $\bb C$ is defined in Eq.\,\pref{Cdef} and $\gamma_i(x)=\gamma_i^\dagger (x)$.  The Hamiltonian then becomes:
\be H=-i\bar v\left[\int_{-\infty}^\infty dx \vec \gamma^T\cdot {d\over dx}\vec \gamma +\vec \gamma^T(0)\bb B\vec \gamma (0)\right]\label{Hgb}
\ee
where $\bb B$ is a $2N$-dimensional real antisymmetric matrix, whose $N(2N-1)$ independent components are linear combinations of the $N(2N-1)$ real components of $\bb M$ and $\Delta$, defined 
in Eq.\,\pref{Hd}.
In this transformed basis, the solutions of the BdG equations have the simple form:
\bea \vec w (x)=
\vec ae^{-ikx}+
\vec be^{ikx}, \qquad  (x>0)
\eea
where $\vec a$ and $\vec b$ are 2$N$-dimensional real vectors, related by the $O(2N)$ ``reflection'' matrix (more appropriately called a transmission matrix after unfolding):
\be \vec {b}=\bb r_M\vec a.\ee
Note that  the physical spatial coordinates are  obtained by  $x\to v_ix/\bar v$, ensuring that the various components of the wave-function 
all have the same energy, $E=v_ik_i=\bar vk$. Next we observe that the boundary term in the Hamiltonian of Eq.\,(\ref{Hgb}) can be eliminated 
by redefining the Majorana fields by:
\be \vec \gamma '(x)\equiv e^{\theta (x)\bb B}\vec \gamma (x)\ee
where $\theta (x)$ is the  step function. It then follows that the reflection matrix is
\be {\bb r}_M=e^{\bb B}.\ee
Since $\bb{B}$ is real and anti-symmetric, $\bb{r}_M$ is an SO(2N) matrix (with determinant $1$).   A second SO(2N) rotation, by $e^{\bb{C}}$,  puts $H_b$ in the form 
\be H_b=t' \gamma \gamma_1''(0).\ee
The corresponding O(2N) reflection matrix, ${\bb r}_0$, is diagonal with entries $(1,-1,1,1,\ldots,1,1)$ and determinant -1. This $\bb{r}_0$ corresponds to the perfect Andreev reflection in the first channel and perfect normal reflect in the other channel. Thus the total reflection matrix is:
\be {\bb r}=e^{\bb B}e^{\bb C}{\bb r}_0.\ee
Therefore $\det ({\bb r})=\det(e^{\bb B})\det(e^{\bb C})\det({\bb r}_0)=-1$; the sign of the determinant of ${\bb r}$ is unaffected by disorder.

\section{Conclusions}
We have discussed the remarkable topological properties of SN junctions from two perspectives, 
one based on a series representation of the $S$-matrix and one based on a low energy 
field theory approach. We have derived conductance bounds for a long 
quantum wire with an ohmic contact to a 3D metal, which depend only on the number 
of channels in the quantum wire (and the topological class of the junction) independent 
of any properties of the 3D metal. 

\acknowledgements 
We than D.~Pikulin for illuminating discussions and C. Beenakker for comments on an earlier 
verison of this manuscript.
This research was supported in part by 
NSERC, CIfAR and the Swiss National Science Foundation.

\appendix

\section{BdG equation and S-matrix}\label{sec:BdG}
\subsection{Bogliubov-de Gennes equations}

We begin by introducing a 4-component spinor of 3-dimensional fermion fields:
\be \Psi (\vec x)\equiv \left(\begin{array}{c} \psi_\uparrow (\vec x)\\ \psi_\downarrow (\vec x) \\ 
\psi^\dagger_\uparrow (\vec x) \\ 
\psi^\dagger_\downarrow (\vec x) \end{array}\right).
\ee
These obey
\bea
\{\Psi_a (\vec x) ,\Psi_b^\dagger (\vec y)\}&=&\delta^3 (\vec x-\vec y),\\  
\{\Psi_a (\vec x) ,\Psi_b (\vec y)\}&=&\tau^x_{ab}\delta (\vec x-\vec y),
\eea
where the indices $a$, $b=1,2,3,4$ and we introduce 4 component Pauli matrices, $\vec \tau$ which act on the particle-hole sectors
\be \tau^x\equiv \left(\begin{array}{cc} \bb 0 &\bb 1\\ \bb 1 & \bb 0\end{array}\right)\ee
et cetera. 
$\bb 0$ and $\bb 1$ are $2\times 2$ zero and unit matrices. In terms of these operators, the second quanitized 
Hamiltonian of Eq.\,(1) can be written
\be H={1\over 2}\int d^3x\Psi^\dagger(\vec x) {\cal H}\Psi (\vec x),\ee
where the Bogliubov-DeGennes (BdG) Hamiltonian is
\be {\cal H}=\left[ {\vec p^2\over 2m}-\mu +V(\vec x)+B(\vec x)\sigma^z\right]\tau^z
+ \alpha\Big(p_x\sigma^y\tau^z-p_y\sigma^x\Big)+\Delta (\vec x)\sigma^y\tau^y.\label{Hq6}\ee
Here $\sigma_i$ and $\tau_i$ for $i=x,y,z$ are pauli matrices in spin and particle-hole bases, respectively
and we have chosen $\Delta (\vec x)$ real and positive for convenience, which can always be done by redefining the phases 
of the fermion fieids.  Note that the single-particle Hamiltonian ${\cal H}$ has the electron-hole symmetry 
\be
\tau^x{\cal H}\tau^x=-{\cal H}^*\label{Heh}
\ee
Without the $p_y\sigma^x$ term (negligible for narrow single-subband wires) the Hamiltonian is real, considering that $p_x=-i\hbar\partial_x$, and has chiral symmetry (class BDI) but in general time-reversal symmetry and spin-symmetry is broken (class D).
We assume that at $x\gg 0$, far from the SN junction on the normal side, the system can be regarded as 
clean so that a reflection matrix can be defined. The asymptotic scattering states can be decomposed into 
$N$ channels of particles and $N$ channels of holes.  These channels in general mix spin components 
due to the spin-orbit interactions. $N$ can be odd or even depending on whether or not the Fermi 
energy lies in between the energy minima of two spin-split channels {[\onlinecite{KAint}]}. 

We introduce a $2N$-component spinor of fermion annihilation and creation operators corresponding to 
the $N$ channels:
\be \Psi (x)\equiv \left(\begin{array}{c}\psi_1(x) \\ 
\psi_2 (x) \\ .\\.\\.\\ \psi_N(x)\\
\psi^\dagger_1(x) \\
\psi^\dagger_2(x) \\.\\.\\.\\ \psi_N^\dagger (x)
\end{array}\right).
\ee 
The most general eigenfunction of energy $E$ has the form in the asymptotic region ($p=\hbar k$)
\be \vec w(x)=\left(\begin{array}{c}
a_{e1}e^{-ik_{e1}x}+b_{e1}e^{ik_{e1}}\\
a_{e2}e^{-ik_{e2}x}+b_{e2}e^{ik_{e2}}\\
.\\.\\.\\
a_{eN}e^{-ik_{eN}x}+b_{eN}e^{ik_{eN}}\\
a_{h1}e^{ik_{h1}x}+b_{h1}e^{-ik_{h1}x}\\
a_{h2}e^{ik_{h2}x}+b_{h2}e^{-ik_{h2}x}\\
.\\.\\.\\
a_{hN}e^{ik_{hN}x}+b_{hN}e^{-ik_{hN}x}
\end{array}\right).\ee
The conserved \emph{quasi-particle current} in the asymptotic region of large positive $x$ is:
\be J=\sum_{j=1}^N \Big[ v_{ej}\Big(\abs{b_{ej}}^2-\abs{a_{ej}}^2\Big)+v_{hj} \Big(\abs{b_{hj}}^2-\abs{a_{hj}}^2\Big)\Big],\label{Jas}
\ee
where $v_{ej}$, $v_{hj}$ is the Fermi velocity for electrons and holes in the $j^{th}$ channel. 
Since we consider $E=0$, $w(x)$ decays exponentially to zero for $x\ll 0$ implying that the current, integrated 
across the cross-section of the quantum wire, obeys
\be J(x)=0,\ \  (\hbox{for all}\ x).\label{J0}\ee
Thus $J$ in Eq.\,(\ref{Jas}) must be zero.

\subsection{S-matrix}
Defining $2N$ component vectors:
\be \vec a\equiv \left(\begin{array}{c} \vec a_e\\ \vec a_h \end{array}\right),\ \ 
\vec b\equiv \left(\begin{array}{c} \vec b_e\\ \vec b_h \end{array}\right),\ee
the right-moving components of the asymptotic wave-function are linearly related to the left-moving components by the $2N\times 2N$ reflection 
matrix $\tilde {\bb r}$ defined by $\vec b=\tilde {\bb r}\vec a$.
Requring Eq.\,(\ref{J0}) to be true for arbitrary incoming wave-function amplitudes, $\vec{a}$ implies the conditions on the 
reflection matrix:
\be \sum_j\tilde {\bb r}^\dagger_{ij}v_j\tilde {\bb r}_{jk}=\delta_{ik}v_i.\ee
where we have defined the $2N$ component vector:
\be\vec v \equiv  \left(\begin{array}{c}\vec v_e\\ \vec v_h \end{array}\right).\\
\ee
It is convenient to define a unitary rescaled reflection matrix:
\be {\bb r}_{ij}\equiv \sqrt{v_i\over v_j}\tilde {\bb r}_{ij}.\label{rtildef}\ee

The electron-hole symmetry property of the BdG Hamiltonian, Eq.\,(\ref{Heh}), 
implies that, at zero energy:
\be
\vec a=\tau^x\vec a^*,
\qquad 
\vec b=\tau^x\vec b^*.
\ee
Thus the zero energy reflection matrix obeys
\be \tilde{\bb r}=\tau^x\tilde{\bb r}^*\tau^x.\ee
Since $v_{ej}=v_{hj}$ at zero energy, the rescaled reflection matrix, ${\bb r}$ also obeys 
this relation, Eq.\,(\ref{reh}).

\section{An alternative proof of conductance range}
We now sketch the details of an alternative proof of the bounds on the conductance for both topological and non-topological junctions and any number of channels, $N$ mentioned earlier in Eq.\,\pref{Gb2}. It is convenient to define
\be \tilde \tau^y \equiv {\bb r}_M^T \tau^y{\bb r}_M\label{tty}\ee
so that Eq.\,(\ref{GM}) becomes:
\be G=\frac{e^2}{h}\left[N-\frac{1}{2}\hbox{Tr}\left(\tilde \tau^y\tau^y\right)\right].\ee
To find the minima and maxima of the conductance, we can treat $\bb{r}_M$ as a variational parameter and look for special $\bb{r}_M$ matrices (and correspondingly $\bar\tau^y$ matrices) for which $G$ has extremal values. Let ${\bb r}_0$ be an $O(2N)$ matrix which gives the maximum or minimum conductance. 
Then $ \hbox{Tr }({\bb r}^T\tau^y{\bb r}\tau^y)$ must be stationary under a small variation
\be {\bb r}(\theta )\equiv e^{\theta \bb{A}}{\bb r}_0\ee
where $\bb A$ is any real antisymmetric matrix. Noting that
\be {d{\bb r}\over d\theta}\bigg|_{\theta =0}=\bb{A}{\bb r}_0,\qquad
{d{\bb r}^T\over d\theta}\bigg|_{\theta =0}=-{\bb r}_0^T\bb{A}\ee
we see that
\be {d\over d\theta}\hbox{Tr}\Big({\bb r}(\theta )^T\tau^y{\bb r}(\theta )\tau^y\Big)\Big|_{\theta =0}=\hbox{Tr }\Big(-{\bb r}_0^T\bb{A}\tau^y{\bb r}_0\tau^y
+{\bb r}_0^T\tau^y\bb{A}{\bb r}_0\tau^y\Big)
=\hbox{Tr }\bb{A}\Big(-\tau^y{\bb r}_0\tau^y{\bb r}_0^T+{\bb r}_0\tau^y{\bb r}_0^T\tau^y\Big)
\ee
Noting that
$({\bb r}_0\tau^y{\bb r}_0^T\tau^y)^T=\tau^y{\bb r}_0\tau^y{\bb r}_0^T$, 
we see that $-\tau^y{\bb r}_0\tau^y{\bb r}_0^T+{\bb r}_0\tau^y{\bb r}_0^T\tau^y$ is anti-symmetric. Therefore the trace of $\bb A$ times this matrix cannot vanish for any anti-symmetric 
matrix $\bb A$ unless 
\be -\tau^y{\bb r}_0\tau^y{\bb r}_0^T+{\bb r}_0\tau^y{\bb r}_0^T\tau^y=[{\bb r}_0\tau^y{\bb r}_0^T,\tau^y]=[\tilde \tau^y,\tau^y]=0.\label{com}\ee\\

Noting that $\tilde \tau^y$ is purely imaginary, anti-symmetric and obeys $(\tilde \tau^y)^2=\bb{1}$, the most general possible $\tilde \tau^y$ 
commuting with $\tau^y$ has the form
\be \tilde \tau^y=i\left(\begin{array}{cc}\bb P&-\bb Q\\
\bb Q&\bb P\\
\end{array}\right)\label{AB}.\ee
Here the  $N\times N$ real matrix $\bb P$ is anti-symmetric and the  $N\times N$ real matrix $\bb Q$ is symmetric with 
\bea \bb{PQ}+\bb{Q P}&=&\bb 0\nonumber \\
\bb Q^2-\bb P^2&=&\bb{1}.\label{BA2}\eea
Letting ${\bb r}_1$ be the $N\times N$ orthogonal matrix which diagonalizes the symmetric matrix $\bb Q$, we can write:
\be \tilde \tau^y=i\left(\begin{array}{cc}{\bb r}_1^T&\bb 0\\\bb 0&{\bb r}_1^T\end{array}\right)\mat{\tilde{\bb{P}}&-\bb\Lambda \\ \bb\Lambda &\tilde{\bb P}}
\left(\begin{array}{cc}{\bb r}_1&\bb 0\\\bb 0&{\bb r}_1\end{array}\right)\label{eq218}
\ee
where $\bb\Lambda$ is diagonal, with entries which are the eigenvalues of  $\bb Q$ and $\tilde{\bb P}\equiv {\bb r}\dn_1\bb{P}{\bb r}_1^T$. Specializing to the case $\bb P=0$, we may use
\be \mat{\bb 0&-\bb\Lambda \\ \bb\Lambda &\bb 0}
=\bb r_D^T\mat{\bb0&-\bb{1}\\ \bb{1} &\bb 0}\bb r_D\ee
where 
\be \bb r_D=\mat{\bb{1}&\bb{0}\\\bb{0}&\bb\Lambda }.\label{eq220}
\ee
The second of Eqs.\,(\ref{BA2}) now implies that all matrix elements  of the diagonal matrix  $\bb\Lambda$ must be $\pm 1$. Furthermore, comparison of Eqs.\,\pref{tty} and (\ref{eq218})-(\ref{eq220}) imply that
\be \det{\bb r}_M=\det \bb r_D\det\left(\begin{array}{cc}{\bb r}_1&\bb 0\\\bb 0&{\bb r}_1\end{array}\right)=\det\bb\Lambda =\prod_{i=1}^N\Lambda_{ii}
\ee
where each $\Lambda_{ii}=\pm 1$. On the other hand
\be G={e^2\over h}[N-\hbox{Tr }\bb \Lambda ].\label{GL}\ee
Noting that all $\Lambda_{ii}=\pm 1$ and that $\det{\bb r}_M=\det\bb\Lambda$ we find
\bea -N&\leq& \hbox{Tr }\bb\Lambda \leq N,\ \  (N\ \hbox{even},\ \hbox{det} \ {\bb r}_M=1)\nonumber \\
  -(N-2)&\leq& \hbox{Tr }\bb\Lambda \leq N,\ \  (N\ \hbox{odd},\ \hbox{det} \ {\bb r}_M=1)\nonumber \\
 -N&\leq& \hbox{Tr }\bb\Lambda \leq N-2,\ \  (N\ \hbox{odd},\ \hbox{det} \ {\bb r}_M=-1)\nonumber \\
   -(N-2)&\leq& \hbox{Tr }\bb\Lambda \leq N-2,\ \  (N\ \hbox{even},\ \hbox{det} \ {\bb r}_M=-1).\label{Lb}
\eea
We see that the maximum and minimum value of $G$ corresponds to all $\Lambda_{ii}$ being $-1$ and $1$ respectively, 
but whether or not this is possible depends on the sign of det ${\bb r}_M$. The maximum value of Tr $\bb \Lambda$ with det ${\bb r}_M=-1$ is $N-2$. 
The minimum value is $-N$ or $-(N-2)$ depending on the parity of $N$ and the sign of det ${\bb r}_M$.
This then leads to Eq. (\ref{Gb2}), plotted in Fig. \ref{fig:Grand}.  We now show that allowing a non-zero matrix $\bb P$ in Eq.\,(\ref{AB}) does not change this range. We again have the two conditions following from $(\tilde \tau^y)^2=\bb{1}$:
\bea \bb\Lambda^2-\tilde{\bb P}^2&=&\bb{1}\nonumber \\
\tilde P_{ij}(\Lambda_{ii}+\Lambda_{jj})&=&0.\eea
Since the square of an  anti-symmetric matrix is negative definite, a non-zero ${\bb P}$, decreases the squares of the eigenvalues of $\bb\Lambda$, 
making it difficult to exceed the upper and lower bounds on $\hbox{Tr }\bb\Lambda$ found above assuming $\bb P=0$. Also note that a matrix 
element $\tilde P_{ij}$ can only be non-zero if $\Lambda_{ii}=-\Lambda_{jj}$.  Thus at least two of the eigenvalues $\Lambda_{ii}$ have opposite sign, 
implying $|\hbox{Tr }\bb\Lambda |\leq N-2$. Thus the bounds of Eq.\,(\ref{Lb}) cannot be exceeded for either parity of $N$ nor either sign of $\det\bb r$.\\

So far we have considered a particular matrix ${\bb r}_M$ which produces a given $\tilde \tau^y$. 
We now prove that {\it all} orthogonal matrices $ {\bb r}_M$ giving the same $\tilde \tau^y$ have the same determinant ($\pm 1$), completing the proof. Noting that
\be {\bb r}_1\tau^y{\bb r}_1^T={\bb r}_2\tau^y{\bb r}_2^T,\ee
implies
\be {\bb r}\tau^y{\bb r}^T=\tau^y\label{rc}\ee
where 
\be {\bb r}\equiv {\bb r}_2^T{\bb r}\dn_1.\ee
We see that the needed result is equivalent to the statement that all orthogonal matrices, ${\bb r}$ obeying Eq.\,(\ref{rc}), or equivalently 
\be [{\bb r},\tau^y]=0,\ee
have determinant +1.
Let $T=i\tau^y$. Consider the eigenvectors of a real matrix, ${\bb r}$, which commutes with $T$. If $u$ is an eigenvector of ${\bb r}$ with 
eigenvalue $\lambda$, then $u^*$ is an eigenvector of ${\bb r}$ with eigenvalue $\lambda^*$.  This implies that, in general, 
the eigenvalues of ${\bb r}$ come in complex conjugate pairs, leading to det ${\bb r}>0$. The only possibility for an unpaired eigenvalue 
is $\lambda$ real and $u$ also real.  (If $\lambda$ were real, and $u$ were not real, then $u^*$ would also be an eigenvector with the same eigenvalue.) 
In this case of a putative unpaired eigenvector, $u$, since $[T,{\bb r}]=0$, it follows that $Tu$ is also an eigenvector of ${\bb r}$ with the same 
real eigenvalue. Of course, since $T$ and $u$ are real, so is $Tu$. But $T^2=-\bb{1}$, so $T^2u=-u$.  Now suppose that $Tu=\lambda 'u$.  If this   
were true $\lambda '$ would have to be real since $u$ and $Tu$ are real. But then we have $T^2u=(\lambda ')^2u=-u$. This implies 
that $\lambda '=\pm i$, which is a contradiction.  So we conclude that $Tu\neq \lambda 'u$ for any $\lambda '$. Therefore 
$Tu$ is another eigenvector of ${\bb r}$, with the same eigenvalue $\lambda$, which is not proportional to $u$.  This contradicts the 
assumption that $u$ was an unpaired eigenvector. More explicitly, we can write the putative unpaired eigenvector as
\be u=\left(\begin{array}{c} v\\w \end{array}\right)\ee
where $v$ and $w$ are $N$-dimensional vectors.  Then
\be Tu=\left(\begin{array}{c} -w\\v \end{array}\right)\ee
$Tu$ can only be proportional to $u$ if $w=\pm iv$ which contradicts the requirement that $u$ is real. \\

It can be seen that {\it all} values of $G$ within the ranges of Eq.\,(\ref{Gb}) can occur. This follows from observing that we can increase 
$G$ from its minimum value in steps of $4e^2/h$ by changing the sign of two of the $\Lambda_{ii}$'s without changing the sign of det ${\bb r}_M$. 
This change in sign of two of the $\Lambda_{ii}$'s corresponds to an $SO(2)$ transformation which can be incorporated into ${\bb r}_1$. 
A continuous set of $SO(2)$ transformations, with rotation angle varying from $0$ to $\pi$, thus covers all values of $G$ in the interval of size $4e^2/h$. 
Eq.\,(\ref{Gb}) is summarized in Fig.\,(\ref{fig:Grand}). \\

\section{Properties of the matrix $\bb r_{eff}$}
In this Appendix we show that the matrix $\bb r_{eff}$ defined in Eq.\,\pref{eqdisorder1} 
is unitary and  particle-hole symmetric if $\bb r$ and $S_1$ matrices are. The particle-hole symmetry of $\bb r_{eff}$ follows from the fact that $\bb r$ and each sub-matrix of $S_1$ separately obey particle-hole symmetry. Therefore,
\be
\tau^x\bb r\dn_{eff}\tau^x=\bb r^*_{eff}.
\ee
Another way to see this would be to use the particle-hole symmetry of $\bb r$ and $S_1$ to represent them as real orthogonal matrices. It follows from Eq.\,\pref{eqdisorder12} that $\bb r_{eff}$ is also real. To establish its orthogonality, it is more convenient to analyse $\bb r\bb r_{eff}$ given by
\be
\bb r\bb r_{eff}=\bb{r}'_{2}+\bb{t}_{2}(\bb{1}-\bb{r}_{2})^{-1}\bb{t}'_{2}.\label{eqdisorder12}
\ee
where the right-hand contains elements of the real and orthogonal matrix $S_2$ defined in Eq.\,\pref{eqS2}, obeying
\bea
\bb r_2^T\bb r_2\dn+\bb t_2^T\bb t_2\dn=\bb 1, \qquad \bb r_2^T\bb t_2'+\bb t_2^T\bb r_2'=\bb 0\nonumber\\
\bb t_2'^T\bb t_2'+\bb r_2'^T\bb r_2'=\bb 1, \qquad \bb t_2'^T\bb r_2+\bb r_2'^T\bb t_2=\bb 0\label{eqS2orth}
\eea
Then we obtain
\be
\bb r^T_{eff}(\bb r^T \bb r\dn)\bb r\dn_{eff}=\bb r_2'^T\bb r_2'+\bb t_2'^T(1-\bb r_2^T)^{-1}\bb t_2^T\bb t_2\dn(1-\bb r_2)^{-1}\bb t_2'+\bb r_2'^T\bb t_2\dn(1-\bb r_2)^{-1}\bb t_2'+\bb t_2'^T(1-\bb r_2^T)^{-1}\bb t_2^T\bb r_2'
\ee
Using the fact that $\bb r^T\bb r=\bb 1$ and the orthogonality conditions \pref{eqS2orth} we can write this expression as
\be
\bb r^T_{eff}\bb r\dn_{eff}=\bb r_2'^T\bb r_2'+\bb t_2'^T \bb X \bb t'_2\label{eqreffX}
\ee
where by $\bb X$ we denote the expression
\bea
\bb X&=&(1-\bb r_2^T)^{-1}\bb t_2^T\bb t_2\dn(1-\bb r_2)^{-1}-\bb r_2(1-\bb r_2)^{-1}-(1-\bb r_2^T)^{-1}\bb r_2^T\\	
&=&(1-\bb r_2^T)^{-1}\Big[(1-\bb r_2^T\bb r_2\dn)-(1-\bb r_2^T)\bb r_2-\bb r_2^T(1-\bb r\dn_2)\Big](1-\bb r_2)^{-1}\\
&=&(1-\bb r_2^T)^{-1}\Big[1+\bb r_2^T\bb r_2\dn)-\bb r_2-\bb r_2^T\Big](1-\bb r_2)^{-1}\\
&=&\bb 1
\eea
Therefore, the right-hand side of Eq.\,\pref{eqreffX} is equal to ${\bb 1}$ owing to orthogonality conditions \pref{eqS2orth} and $\bb r^T_{eff}\bb r\dn_{eff}=\bb 1$.


\begin{thebibliography}{50}
\bibitem{odd} Turning off the pairing term in the superconducting region, there must be an odd number of active spin-split channels 
if the system is in the topological phase. However, by changing parameters such as the magnetic field 
at the junction it is possible to have an even number of active channels in the normal region. An example of this 
is discussed in [\onlinecite{KAint}], where mechanisms for abruptly changing the magnetic field are discussed.
\end{thebibliography}
\end{document}